\documentclass[12pt]{article}
\usepackage{amsfonts,amssymb,amsthm,eucal,amsmath,verbatim,color,bbm}
\usepackage{graphicx}
\allowdisplaybreaks
\newtheorem{thm}{Theorem}

\newtheorem{prop}[thm]{Proposition}

\newcommand{\R}{\mathbb{R}}
\newcommand{\Z}{\mathbb{Z}}

\newcommand{\E}{\mathbb{E}}

\newcommand{\inprod}[2]{\left\langle #1, #2 \right\rangle}

\newcommand{\ee}{\mathbb{E}}

\newcommand{\s}{\mathbb{S}}

\newcommand{\1}{\mathbbm{1}}

\begin{document}

\title
{Critical behavior of mean-field XY and related models}
\author{Kay Kirkpatrick and Tayyab Nawaz\thanks{Both authors partially supported by NSF CAREER award DMS-1254791, and NSF grant 0932078 000 while in residence at MSRI during Fall 2015.}  \\ \\
Department of Mathematics, University of Illinois at Urbana-Champaign\\
1409 W. Green Street, Urbana, IL 61801, USA}

\maketitle


\begin{abstract} 
We discuss spin models on complete graphs in the mean-field (infinite-vertex) limit, especially the classical XY model, the Toy model of the Higgs sector, and related generalizations. We present a number of results coming from the theory of large deviations and Stein's method, in particular, Cram\'er and Sanov-type results, limit theorems with rates of convergence, and phase transition behavior for these models.
\end{abstract} 

\section{Introduction}

We use mean-field theory to approximate a challenging problem and to study a many-body problem by converting it into a one-body problem. We survey a number of results obtained recently using the theory of large deviations along with Stein's method-type limit theorems to describe the asymptotic behavior of the $O(N)$ spin models such as the $N=1$ Curie-Weiss model, the $N=2$ model called the XY model, the $N=3$ Heisenberg model, and the $N=4$ Toy model of the Higgs sector \cite{CS,KM,KN}. We present these results mostly without proofs. In this section, we describe the mean-field XY model and the history, including the 2D XY model (which is currently intractable). In the next section we describe the asymptotic behavior of the XY model; in the last section, the behavior of its generalizations to $N$-dimensional spins.

The XY model on a complete graph $K_n$ with $n$ vertices in the absence of an external field is defined as follows: there is a circular spin $\sigma_i \in \s^1$ at each site $i \in {1,2,...,n}$. The configuration space of the XY model is $\Omega_n = (\s^1)^n$ where each microstate is $\sigma = {(\sigma_1,\sigma_2,...,\sigma_n)}$. For the higher 
$O(N)$ spin models, we simply replace $\s^1$ by  $\s^{N-1}$, and in all cases the Hamiltonian energy is defined by

$$H_n(\sigma) = - \sum_{i,j} J_{i,j}\inprod{\sigma_i}{\sigma_j}.$$
\
The mean-field interaction for the XY and $O(N)$ models on the complete graph is defined by $J_{i,j} = \frac{1}{2n}$ for all $i,j$.

The simplest spin model is the Ising model, with one-dimensional $\pm1$ spins, a model that is  used extensively in statistical mechanics, invented by Ernest Ising while working with his advisor Wilhelm Lenz \cite{EZ,SGB}. The one-dimensional Ising model has no phase transition, but there is a phase transition on an infinite two-dimensional lattice. 
The mean-field Ising model, or Curie-Weiss model, describes the Ising model well for higher dimensions, and the magnetization (average spin) in this model has a Gaussian law away from the critical temperature and a non-Gaussian law at the critical temperature \cite{EN}. 
Recently, Chatterjee and Shao \cite{CS} proved that the total spin in the Curie-Weiss model at the critical temperature satisfies a Berry--Esseen type error bound in this non-Gaussian limit.

The XY model, with two-dimensional circular spins, models superconductors and is interesting but challenging to study its phase transition rigorously  \cite{PWA}. On a lattice of two spatial dimensions, such a continuous circular symmetry cannot be broken at any finite temperature \cite{MW}. Thus the 2D XY model cannot have an ordered phase at low temperature quite like the Ising model, and it has a phase transition that is quite different from the Ising model
\cite{MMA,SHE}. Instead, the 2D XY model exhibits the peculiar Kosterlitz-Thouless (KT) transition,
a phase transition of infinite order and the subject of a Nobel prize. Above the transition temperature $T_{KT}$ correlations between spins decay exponentially. At low temperatures, the system does not have any long-range order as the ground state is unstable, but there is a low-temperature quasi-ordered phase with a correlation function that decreases with the distance like a power, which depends on the temperature \cite{KT}.

Because the 2D XY model is so challenging, we study the mean-field classical XY model instead, which can be viewed as the large-dimensional ($d \to \infty$) limit of the nearest-neighbor model on $\Z^d$, with spins in $\s^{1}$, and with critical inverse temperature $\beta_c=2$ \cite{KS}. Furthermore,  the large-dimensional limit approximates high-dimensional models nicely since below the critical temperature, the average spin goes to zero for all $d$, and above the critical temperature, the total spin has a non-zero limit as $d \to \infty$.

In the next section we will examine the XY model in detail, while section 3 deals with extensions to higher spin dimensions. 

\section{The mean-field XY model and asymptotic results}

\begin{figure}[h]
	\centering
	\includegraphics[width=4in]{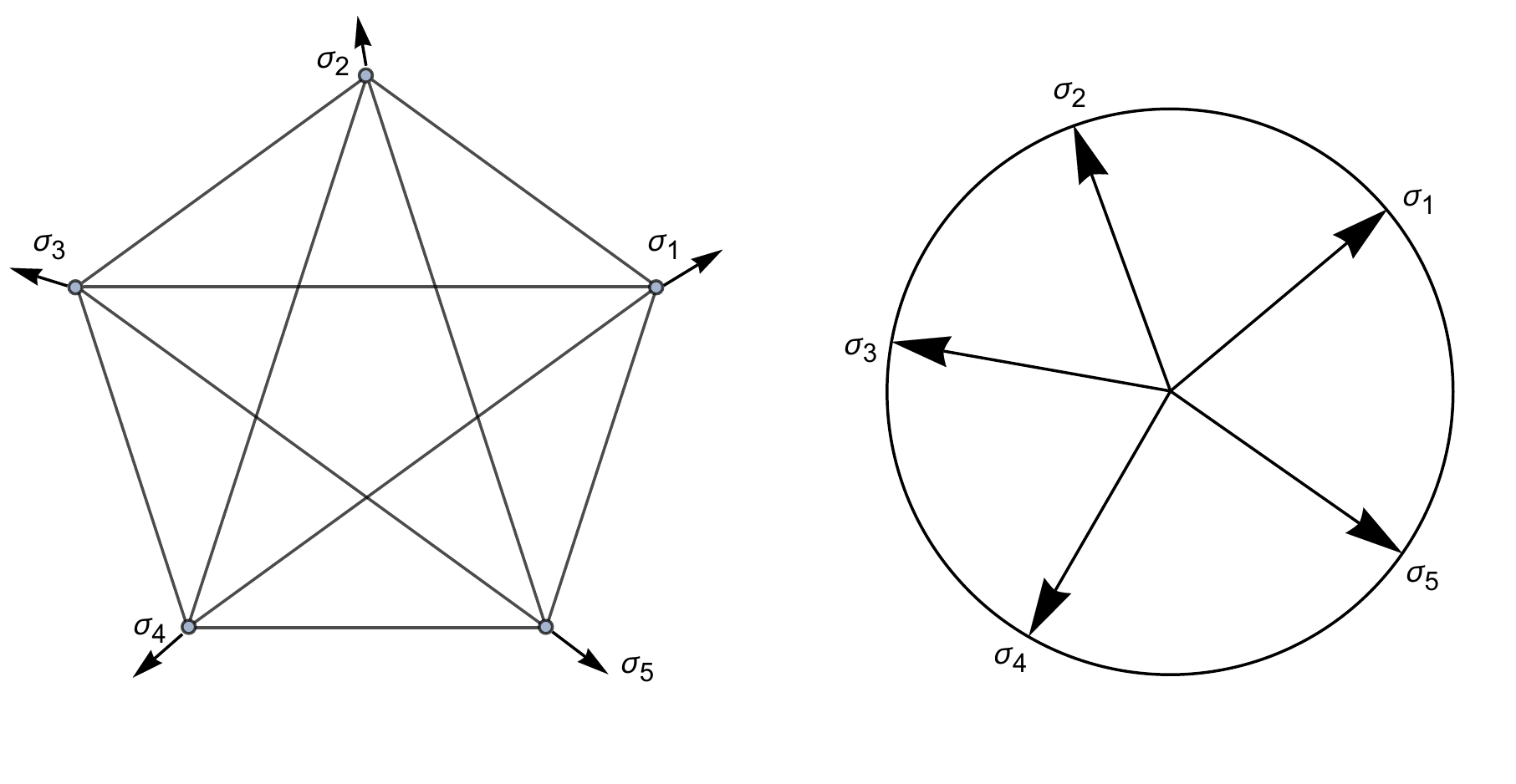}
	\caption{Left: The classical mean-field XY Model on the complete graph $K_5$ with five sample spin vectors. Right: The projection of the same spin vectors from $K_5$ onto $\s^1$.}
	\label{fig:1}
\end{figure}

We consider the isotropic mean-field classical XY model on a finite complete graph $K_n$ with $n$ vertices. That is, at each site $i \in K_n$ of the graph is a spin living in $\Omega = \s^1$, so the state space is $\Omega_n = (\s^1)^n$. See Fig. \ref{fig:1} for a picture of the XY model on 5 vertices.
The corresponding mean-field Hamiltonian energy $H_n : \Omega_n \to \R$ is given by: 
\[ H_n (\sigma) :=  -\frac{1}{2n} \sum_{i,j=1}^n\inprod{\sigma_i}{\sigma_j} = -\frac{1}{2n} \sum_{i,j} \cos (\theta_i - \theta_j), \]
where $\theta_i$ is the angle that the $i$-th spin makes with respect to some axis.
The corresponding Gibbs measure is the probability measure 
$P_{n,\beta}$ on $\Omega_n$ with density function: 
\begin{equation} \label{Gibbs-measure}
f(\sigma):=\frac{1}{Z(\beta)}e^{-\beta H_n(\sigma)}.
\end{equation}
where $Z$ is the normalizing constant, also known as the partition function, which encodes the statistical properties of the model such as free energy and magnetization. Note that Gibbs measure here is a normalization of the Boltzmann distribution, and that the inverse temperature $\beta$ is equal to $(k_B T)^{-1}$, where $k_B$ is the Boltzmann constant and $T$ is the temperature of the system. We can understand the structural behavior of the spin vectors' distribution by studying extreme cases for the inverse temperature $\beta$ as follows:
\begin{itemize}
\item At high temperature, from equation (\ref{Gibbs-measure}) we can predict that the Gibbs measure is uniform.
\item At low temperature, again from equation (\ref{Gibbs-measure}) we can predict that the Gibbs measure decays quickly, and the spin vector distribution prefers the lowest-energy ground state.
\end{itemize}
The most likely physical system states corresponding to the Gibbs measure are called the canonical macrostates. We will consider the random measure of the spins $\{\sigma_i\},$ defined by $\mu_{n,\sigma}:=\frac{1}{n}\sum_{i=1}^n\delta_{\sigma_i}$ on $\s^1$ and study the total empirical spin, defined by
\[ S_n (\sigma) := \sum_{i=1}^n \sigma_i. \]

The relative entropy of a probability measure $\nu$ on $\s^1$, with respect to 
the uniform probability measure $\mu$ is defined by
\begin{equation} \label{Relative-Entropy}
H(\nu\mid\mu):=\begin{cases}\int_{\s^1}f\log(f)d\mu& if \,\, f:=\frac{d\nu}{d\mu} 
\,\,exists;\\\infty&otherwise.\end{cases}
\end{equation}

\subsection{LDPs, free energy, and macrostates for the XY model}

Let $M_1(\s^1)$ represent the probability measures on $\s^1$ with the weak-$*$ topology. We are interested in analyzing the total spin, $S_n := \sum_{i=1}^{n} \sigma_i$, as a function of the inverse temperature $\beta$ in the Gibbs measures. This leads us to consider large deviation
principles (LDPs) for the $ \mu_{n,\sigma}$, and then we can rewrite the free energy more explicitly to describe the phase transition at $\beta = 2$.  Part of Theorem~\ref{LDP1} ($\beta = 0$) is a special case of  Sanov's theorem, while the other cases ($\beta > 0$) follow from an abstract result of Ellis, Haven, and Turkington (\cite{EHT}, Theorem 2.5).

\begin{thm}\label{LDP1}
 If $P_n$ is the $n$-fold product of uniform measure on  $\s^1$ and $\mu_{n,\sigma}$ is the random measure as defined above. 
For $\Gamma \subset M_1(\s^1)$, the  $\mu_{n,\sigma}$ satisfy an LDP with rate function
\begin{equation}\label{beta_rf}
I_\beta(\nu):=H(\nu\mid\mu)-
\frac{\beta}{2}\left|\int_{\s^1}xd\nu(x)\right|^2-\varphi(\beta),
\end{equation}
where the free energy is given by 
\begin{equation}\label{free-energyxy}\varphi(\beta)=
\inf_{\nu\in M_1(\s^1)}\left[ H(\nu\mid\mu)-
\frac{\beta}{2}\left|\int_{\s^1}xd\nu(x)\right|^2\right].\end{equation}
For fixed $\beta \ge 0$, every subsequence of $P_{n,\beta}\left[\mu_{n,\sigma}\in\cdot\right]$ converges weakly to a probability measure on $M_1(\s^1)$ concentrated on  the canonical macrostates $\mathcal{E}_\beta:= \{\nu: I_\beta(\nu)=0\}$, i.e., the zeros of the rate function.
\end{thm}

For $\beta = 0$, the relative entropy $H(\cdot\mid\mu)$ achieves its minima of $0$
only for the uniform measure $\mu$, implying that the canonical macrostate is disordered.
For $\beta > 0,$ canonical macrostates are defined abstractly through zeros of the rate function (\ref{beta_rf}), and later Theorem \ref{freeenergyresults} will describe the macrostates explicitly. 

The free energy given by (\ref{free-energyxy}) can be transformed into the following more explicit form.

\begin{thm}\label{freeenergy} (Kirkpatrick-Nawaz \cite{KN})
 The free energy $\varphi$ has the formula:
\[ \varphi(\beta) = \begin{cases} 0, \quad \quad \quad \quad\; \text{ if } \beta <2, \\ \Phi_\beta(g^{-1}(\beta)), \text{ if } \beta\ge 2, \end{cases}\] 
where $I_i$ below are modified Bessel functions of first kind and $\Phi_\beta$ is the functional defined by: \begin{equation}\label{Phi}
\Phi_\beta(r) := r \frac {I_1(r)}{I_0(r)}-\log\left[ I_0(r) \right]-\frac{\beta}{2}\left(\frac {I_1(r)}{I_0(r)}\right)^2,
\end{equation}
 and \[ g(r) := r \frac {I_0(r)}{I_1(r)} .\]
 Here the phase transition is continuous as the function $\varphi$ and its derivative $\varphi'$ are continuous at the critical threshold $\beta = \beta_c = 2$.
\end{thm}

The magnetization for the classical XY model can be obtained by differentiating the partition function: 
\[|m| = \left| \ee\left[\frac1n \sum_i \sigma_i\right]\right|  =  \left| \ee\left[\frac1n S_n\right] \right| = \frac{I_1(r)}{I_0(r)}\]

From the free energy we can precisely explain the phase transition at $\beta = 2$. For $0 \le \beta \leq 2$, we have a unique global minimum for the free energy at the origin with a zero magnetization. For $\beta \geq 2$, we have a unique global minimum for a positive radius.  
 
Let $\{\sigma_i\}_{i=1}^n$ be i.i.d.\ uniform random points on $\s^1\subseteq\R^2$. We have the following Cram\'er-type LDP for the average spin.

\begin{thm}\label{spinLDPbeta}  (Kirkpatrick-Nawaz \cite{KN})
Let $P_{n,\beta}$ be the Gibbs measure defined above \eqref{Gibbs-measure}. Then for $\beta \ge 0$, the average spin
  $M_n=M_n(\sigma):=\frac1n \sum_{i=1}^n \sigma_i$ satisfies an LDP with rate function $I_\beta (x) = \Phi_\beta (r )$:
  \[P_{n,\beta} \left( M_n \simeq x \right) \simeq e^{-n \Phi_\beta(r)},\]
  where $\Phi_\beta$ is given by (\ref{Phi}) and $r = |x|$.
\end{thm}

For an explicit representation of $\mathcal{E}_\beta$, we note from (\ref{Relative-Entropy}) that the relative entropy depends only on the distribution of $f$.  By Fubini's theorem 
\[\int f\log(f)d\mu=\int_0^\infty \mu\big[f\log(f)>t\big]dt-\int_0^\infty \mu\big[f\log(f)<-t\big]dt.\]
This implies that for a fixed $f$, the quantity $\left|\int xd\nu(x)\right|$ is maximized for corresponding densities which are 
symmetric about a fixed pole and decreasing as the distance from the pole increases.  Using this reasoning, consider a density $f$ that is symmetric about the north pole and decreasing away from the pole i.e., $\nu_f$ is the measure with density $f(x,y)=f(y)$ which is increasing in $y$. Then
\begin{equation*}\begin{split}
H(\nu_f\mid\mu)&=\frac{1}{2\pi} \int_{\s^1} f(x,y) \log[f(x,y)] dx dy \\&=\frac{1}{\pi} \int_0^{2\pi}\int_0^\pi f(\cos(\theta))\log[
f(\cos(\theta))]d\theta d\varphi \\& = \frac{1}{\pi} \int_{-1}^{1} \frac{f(y) \log[f(y)]}{\sqrt{1-y^2}} dy.
\end{split}\end{equation*}
Similarly,
\[\int_{\s^1} x d\nu_f(x)=\frac{1}{\pi} \begin{bmatrix}0\\1\end{bmatrix}\int_{-1}^1
\frac{y f(y)}{\sqrt{1-y^2}} dy.\]
Therefore, our minimization problem is reduced to minimizing the following functional
\[ \frac{1}{\pi} \int_{-1}^{1} \frac{f(y) \log[f(y)]}{\sqrt{1-y^2}} dy-\frac{\beta}{2}\left( \frac{1}{\pi} \int_{-1}^1
\frac{y f(y)}{\sqrt{1-y^2}} dy\right)^2\]
over $f:[-1,1]\to\R_+$ such that $f$ is increasing and $\frac{1}{\pi} \int_{-1}^{1}\frac{f(y)}{\sqrt{1-y^2}} dy =1$.  
We can rewrite the first term of the last expression to see that it involves the usual entropy $S(f) = \int f \log(f)$: 
 \[\frac{1}{\pi} \int_{-1}^{1} \frac{f(y) \log[f(y)]}{\sqrt{1-y^2}} dy=\frac{1}{\pi}\int_{-1}^1 \frac{f(y)}{\sqrt{1-y^2}} \log\left[\frac{f(y)}{\pi}\right]dy+\log(\pi)=-S\left(\frac{f}{\pi}\right)+\log(\pi).\]

Now for $\left|\int xd\nu(x)\right| = c $ $\in[0,1]$, using constrained entropy maximization (see Theorem 12.1.1 from \cite{CT}),
we will minimize $\frac{1}{\pi} \int_{-1}^{1}\frac{y f(y)}{\sqrt{1-y^2}} dy$,  that is, maximize $S(f/\pi)$,
over the $\nu\in M_1(\s^1)$ corresponding to this $c$.
\begin{prop}\label{entropy_max} (Kirkpatrick-Nawaz \cite{KN})
Consider a set of functions $f:[-1,1]\to\R_+$, with weight function $w(y) = \frac{1}{\sqrt{1-y^2}}$, such that $\int_{-1}^1 f(y) w(y) dy=1$, and $\left|\int_{-1}^1 yf(y)w(y)dy\right|=c.$ i.e., weighted integral of $f$ is 1 while first weighted moment is bounded. Then the exponential function $f^*(y)=\pi a e^{b y}$ uniquely maximizes $S(f/\pi)$ over the 
densities satisfying these conditions.
\end{prop}

For $c\in[0,1]$, observe that $f^*$ increasing gives all $b \in[0,\infty)$.
Now for $b\in[0,\infty)$, our functional minimization reduces to the following one dimensional function:
\begin{equation}\begin{split}\label{tbm}
\frac{1}{\pi} \int_{-1}^{1} \frac{f(y) \log[f(y)]}{\sqrt{1-y^2}} dy-\frac{\beta}{2}c^2&= b \frac {I_1(b)}{I_0(b)}-\log\left[ I_0(b) \right]-\frac{\beta}{2}\left(\frac {I_1(b)}{I_0(b)}\right)^2\\&=: \Phi_\beta(b).
\end{split}\end{equation}
   
The following theorem, a special case proved using the calculus of variations in \cite{KN}, describes the canonical macrostates: \newline


\begin{thm}\label{freeenergyresults} (Kirkpatrick-Nawaz \cite{KN})

\begin{enumerate}
\item For $\beta\le 2$, $\inf_{b\ge 0}\left\{\Phi_\beta(b)
\right\}=0$ is achieved for $b=0$ and the corresponding $a=1$, so that the minimizing function $f^* = 1$ and therefore the only canonical macrostate is the uniform distribution $\mu$.
\item For $\beta>2$, $\inf_{b\ge 0}\left\{\Phi_\beta(b)
\right\} = \Phi_\beta(g^{-1}(\beta))$, where $b=g^{-1}(\beta)$ is the unique strictly positive solution to $g(b ) = \beta$ where 
\[g(b)=b \frac{I_0(b)}{I_1(b)},\] 
$a = \frac{1}{\pi I_0(b)}$and $\lim_{\beta \downarrow 2} \inf_{b\ge 0}\left\{\Phi_\beta(b)
\right\}=0$. In this case, the canonical macrostates are given by
$\mathcal{E}_\beta = \{\nu_{f,x}\}_{x\in\s^1},$ where $\nu_{f,x}$ is the measure that is the rotation of $\nu_f$ from north pole to x-direction, which is symmetric about the north pole with density $f:[-1,1]\to\R$ given by $f(y) = \pi ae^{b y}$ with $a$ and $b$ as above.
\end{enumerate}
\end{thm}

\begin{figure}[h]
\centering
\includegraphics[width=4in]{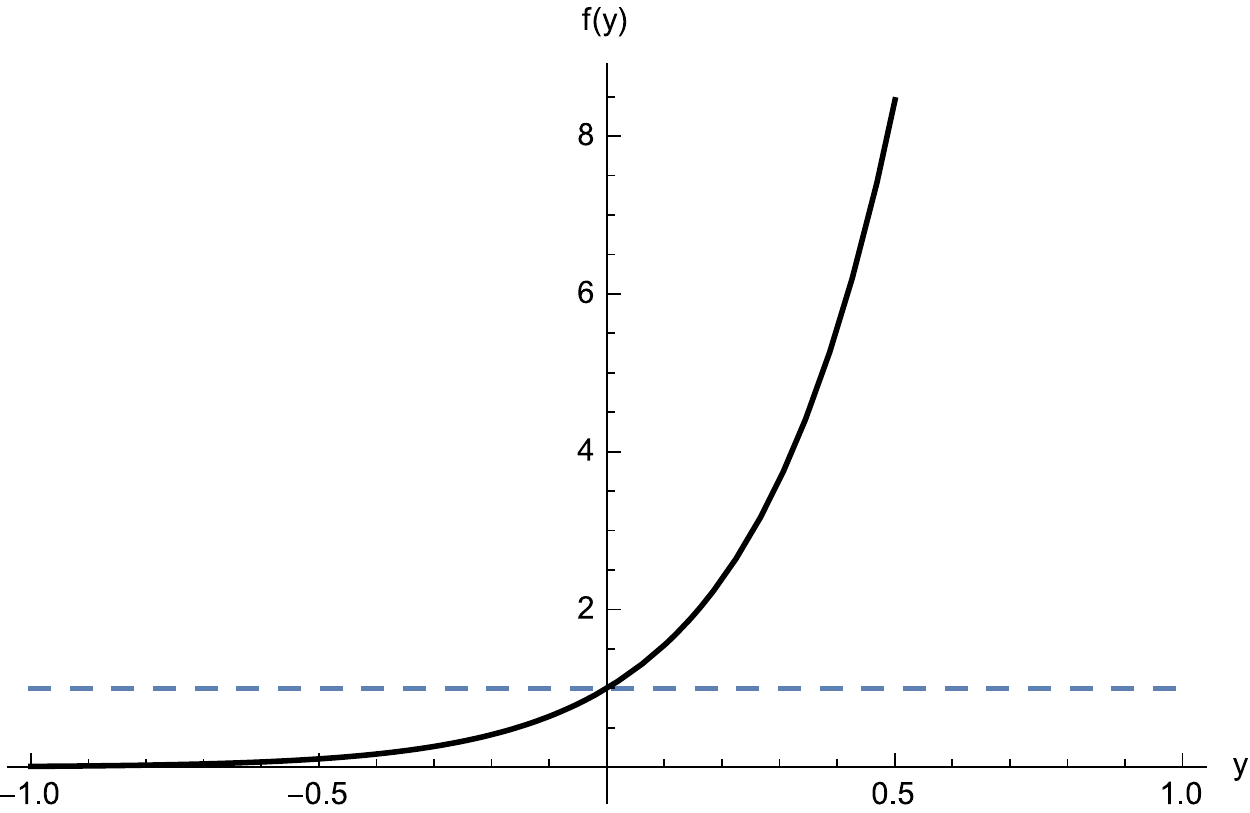}
\caption{Cross-sections of two canonical macrostates: For $\beta \le 2$ (the disordered regime), we have the uniform distribution $f(y) =1$ as the dotted line; for $\beta = 5 > \beta_c = 2$ (the ordered regime), we have plotted the cross-section of the distribution $\nu_f$, given by $f(x,y) = f(y) = \frac{e^{b y}}{I_0(y)}$, showing that the spins point predominantly near the north-pole direction.}
\label{fig:2}
\end{figure}

We can also visualize the Gibbs measure corresponding to subcritical or supercritical cases as shown in Fig. \ref{fig:2}.

\subsection{Limit theorems for the total spin in the XY model}
Next we understand the asymptotics for the total spin of the mean-field XY model, in different regimes across the phase transition, describing the central and non-central limit theorems for each phase. 

In the high temperature regime $\left(0 \le \beta < 2\right)$, the average spin (magnetization) of the system goes to zero with increasing number of spins $n \to \infty$, and we have a multivariate central limit theorem with a rate of convergence in Theorem \ref{subcrit_limit}. The main idea is to use Stein's method  \cite{KM,St,Me} with the exchangeable pair $(W_n,W_n')$ from the Gibbs sampling approach: our random variable representing the rescaled total spin of the original configuration is 
$$W_n:=\sqrt{\frac{2-\beta}{n}}\sum_{i=1}^n\sigma_i,$$ while the random variable representing the rescaled total spin of the new configuration, with $I \in \{1, \dots, n\}$ chosen uniformly at random, is 
$$W_n':=W_n(\sigma')=
W_n-\sqrt{\frac{2-\beta}{n}} \sigma_I+\sqrt{\frac{2-\beta}{n}}\sigma_I'.$$

\begin{thm} \label{subcrit_limit} (Kirkpatrick-Nawaz \cite{KN}) In the high temperature regime $0<\beta<2$, if $W_n$ is defined as above, $Z$ is a standard normal random variable in $\R^2$, $c_\beta$ is a function depending on $\beta$ only, $L(g)$ is the modulus of uniform continuity of $g$, and  $M(g)$ is the maximum operator norm of the Hessian of $g$, then we have: 
\[\sup_{g:L(g),M(g)\le 1}|\E g(W_n)-\E g(Z)|\le\frac{c_\beta}{\sqrt{n}}\]
\end{thm}

The proof of Theorem \ref{subcrit_limit} proceeds in several steps, as a special case of \cite{KN}: first we use the fact that the density of the Gibbs measure is rotationally invariant to conclude that each spin has a uniform marginal distribution. We obtain the complete asymptotic behavior of the total spin using the rotational invariance of the total spin, a strategy adapted from \cite{KM}. We calculate the variance of the total spin to arrive at the proper scaling for defining the exchangable pair and use the pair to derive expressions and bounds for the linear factor $\Lambda$ appearing in the conditional expectation and the remainder terms $R$ and $R'$ \cite{KM, KN, Me}. The rest follows from a theorem of Meckes \cite{Me}. 

As the temperature decreases to zero, the spins start aligning. For smaller values of  $\beta > 2$, the spins vectors are aligned weakly, while for larger $\beta$, this alignment is strong. For any $\beta > 2$, because of the large deviation principle in Theorem \ref{spinLDPbeta}, we have that $|\sum \sigma_j |$ is close to $b n/\beta$ with high probability, if $b$ is the minimizer in $\Phi_\beta$. And due to the circular symmetry, all points on the circle of radius $b n/\beta$ are equally likely. With this reasoning, similar to \cite{KM}, it is natural to consider the random variable representing the fluctuations of squared-length of total spin, i.e.,

 \begin{equation}\label{W-def}
W_n:=\sqrt{n}\left[\frac{\beta^2}{n^2b^2}\left|\sum_{j=1}^n\sigma_j 
\right|^2-1\right].\end{equation}
\smallskip
Our multivariate central limit theorem in the low temperature (ordered) regime is as follows:
\begin{thm}\label{T:supcrit_CLT} (Kirkpatrick-Nawaz \cite{KN})
If $\beta>2$ and $b$ is the solution of $b= \beta f(b) := \beta \frac {I_1(b)}{I_0(b)},$ and $W_n$ is as defined above in \eqref{W-def}, and if $Z$ is a centered normal random variable with variance $V$, where 
\[V=\frac{4\beta^2}{\left(1-\beta f'(b)\right)b^2}
\left[1-\frac{1}{b}\frac{I_{1}(b)}{I_{0}(b)}-\left(\frac{I_{1}(b)}{I_{0}(b)}\right)^2\right],\]
then there exists $c_\beta$, depending only on
$\beta$, such that 
 then
\[d_{BL}(W_n,Z)\le c_\beta\left(\frac{\log(n)}{n}\right)^{1/4}.\]
where $d_{BL}(X,Y)$ is the bounded Lipschitz distance between random
variables $X$ and $Y$.
\end{thm}
Again the proof of Theorem \ref{T:supcrit_CLT} follows from a univariate analogue of the abstract normal approximation of Stein \cite{St}, and relies on conditional moment bounds. The fact that the variance is positive was proved by Amos \cite{AMOS} while deriving the improved bounds on the ratio of Bessel functions.

At the critical temperature $\beta_c=2$, we will consider the random variable 
\begin{equation}\label{Wsl-def}
W_n:=\frac{c}{n^{3/2}}\sum_{i,j=1}^n\inprod{\sigma_i}{\sigma_j},\end{equation}
and make an exchangeable pair $(W_n,W_n')$ using Glauber dynamics. Using symmetry of the total spin and Stein's method similar to \cite {CS,KM}, we will obtain critical limiting density function $p$ as defined below. 

\begin{thm}\label{T:limit_crit} (Kirkpatrick-Nawaz \cite{KN})
For the critical inverse temperature $\beta=2$, if $W_n$ is as defined above in \eqref{Wsl-def}, and $X$ is the random variable with the density
 \[p(t)=\begin{cases} \frac{1}{Z}e^{-t^2/64}&t\ge
   0,\\0&t<0,\end{cases}\]
where  $Z$ is normalizing constant, then there exists a universal constant $C$ such that 
\[\sup_{\substack{\|h\|_\infty\le 1, \,\|h'\|_\infty\le 1\\\|h''\|_\infty\le
1}}\big|\E h(W_n)-\E h(X)\big|\le\frac{C\log(n)}{\sqrt{n}}.\]
\end{thm}

The proof of the limit theorem for the critical temperature is essentially via the ``density approach'' to Stein's method introduced by Stein, Diaconis, Holmes, and Reinert \cite{SDHR}. Recenlty, also  Chatterjee and Shao \cite{CS} have applied this approach to the total spin of the mean-field Ising model, i.e., the Curie-Weiss model. 

We note that these limit theorems with explicit rates of convergence can be generalized to high-dimensional spins, but we will omit those technicalities in the following section.

\section{High-dimensional spin $O(N)$ models }

We can use similar methods to extend our results for two-dimensional spin classical XY model to classical $O(N)$ models, or $N$-vector models. In this general case, with spins in $\s^{N-1} \subset \R^N$, the critical inverse temperature is $\beta_c=N$ \cite{KS,KN}. The $N$-vector models on a complete graph $K_n$ have the Hamiltonian:
 \begin{equation}\label{N-dimensionH}
H_n (\sigma) :=  -\frac{1}{2n} \sum_{i,j=1}^n\inprod{\sigma_i}{\sigma_j}.\end{equation} We present results about the magnetization, free energy, and critical behavior in the $O(N)$ models. It is important to note that we divide our asymptotic analysis into two cases: if $N$ an even positive integer, we have modified Bessel functions of first kind with order $\nu = N/2$ and $\nu - 1$, while for $N$ odd, we have hyperbolic functions arising from the half-integer order Bessel functions. 

\subsection{The magnetization in $O(N)$ models}

Similar to the classical XY model, we can calculate the magnetization of the classical $N$-vector unit hyperspherical model using the conditional density, from the conditional expectations, and it turns out to be a ratio of modified Bessel function of first kind:
\begin{thm}\label{magnetization} (Kirkpatrick-Nawaz \cite{KN})
Consider the $O(N)$ model with the above Hamiltonian (\ref{N-dimensionH}), with $N$ representing the dimension of the spin $\sigma_i \in \s^{N-1}$. Then  on the complete graph $K_n$ the $O(N)$ magnetization $M_{N,n}= \sum_{i=1}^n \sigma_i$ has the following mean-field limit:
\[\left| M_N\right|  = \frac{I_\frac{N}{2}(r)}{I_{\frac{N}{2}-1}(r)}\]
\end{thm}

\begin{figure}[h]
\centering
\includegraphics[width=4in]{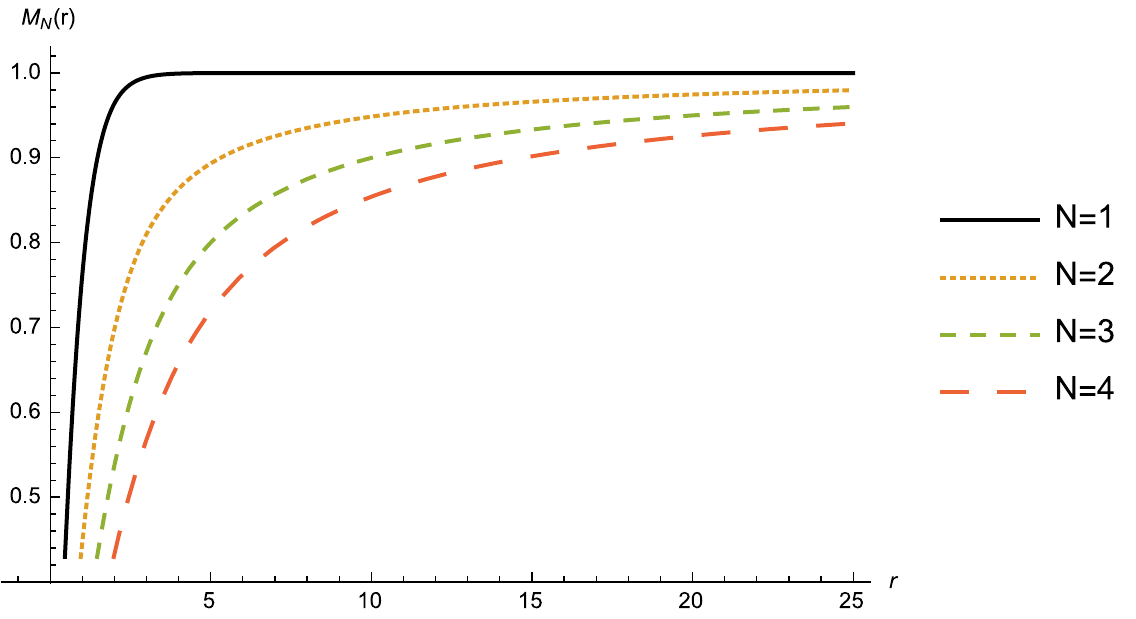}
\caption{Graph of magnetization limits $|M_N|$ for $N$-vector models, $1 \le N \le 4.$ For the mean-field Ising model, $M_1 = tanh(x)$, for the mean-field XY model $\left| M_2\right|  = \frac{I_1(r)}{I_0(r)}$, for the mean-field Heisenberg model $\left| M_3\right|  = coth(r)-\frac{1}{r}$, and for the mean-field Toy model of the Higgs sector, $\left| M_4\right|  = \frac{I_2(r)}{I_1(r)}$. Here $r$ and $\beta$ are related by the formula
$g_N(r) := r \frac{ I_{\frac{N}{2}-1}(r) }{ I_\frac{N}{2}(r)} = \beta$ }
\label{fig:3}
\end{figure}

From Fig \ref{fig:3}, we can observe that low-dimensional spin models can be magnetized easier in some sense, and as the spin gets higher dimensional, it takes more energy to magnetize the physical system.

\subsection{The rate function and free energy in $O(N)$ models}

Next we will present rate functions for large deviation principles similar to Theorems \ref{LDP1}\&\ref{spinLDPbeta}, the first of which is the relative entropy for the $N$-vector model given by an abstract formula similar to before:
\[I_{\beta,N}(\nu):=H(\nu\mid\mu)-
\frac{\beta}{2}\left|\int_{\s^{N-1}}xd\nu(x)\right|^2-\varphi(\beta)\]
where $H(\nu\mid\mu)$ is the relative entropy (\ref{Relative-Entropy}) and $\varphi_N$ is the free energy defined abstractly as before:
\begin{equation}\label{free-energy}\varphi_N(\beta)=
\inf_{\nu\in M_1(\s^{N-1})}\left[ H(\nu\mid\mu)-
\frac{\beta}{2}\left|\int_{\s^{N-1}}xd\nu(x)\right|^2\right].\end{equation} We can calculate the minima in the expression of this rate function and verify that in the subcritical regime ($\beta < N$) there is a unique minimum, while in the supercritical regime there is a family of minima parametrized by $\s^{N-1}$. The free energy given by (\ref{free-energy}) can be written in the following more explicit form using a method like the one in the previous section. In particular, we have a Cram\'er-type LDP for the average spin $M_n := \frac1n \sum_{i=1}^n \sigma_i \in \R^N$,
with rate function $I_{\beta,N}(x) = \Phi_{\beta,N}(r)$, defined below for $\beta \ge 0$ and $r = |x|$.
\begin{figure}[h]
\centering
\includegraphics[width=4in]{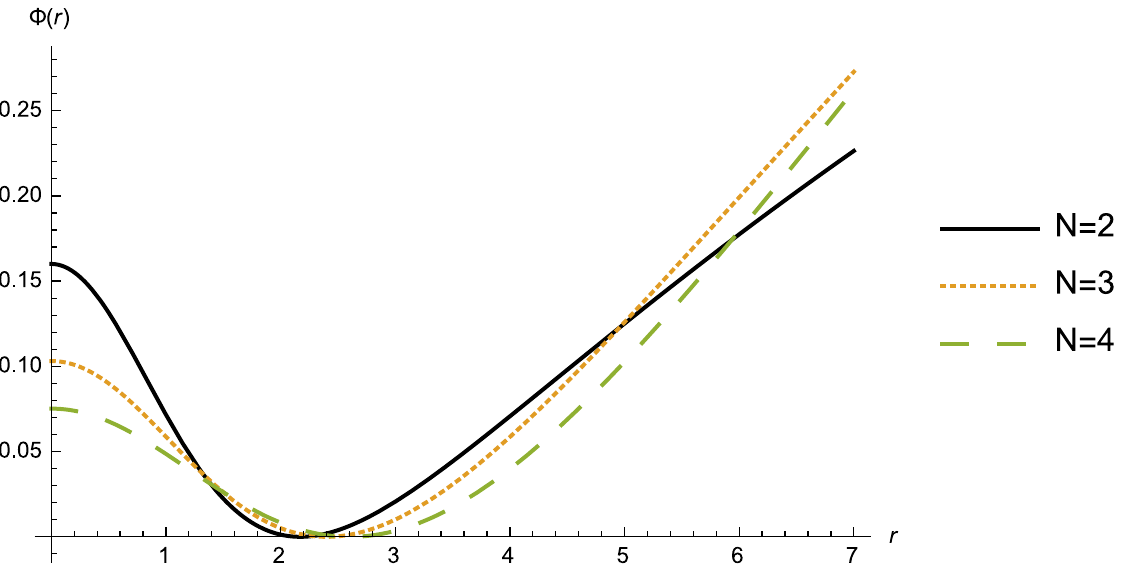}
\caption{Graph of the rate function $I_{\beta,N}(x) = \Phi_{\beta,N}(r)$ in the supercritical regime ($\beta = N+1$) for $2 \le N \le 4$, which has minimum at radius $g^{-1}_N(\beta) = r$.}
\label{fig:4}
\end{figure}

\begin{thm}\label{freeenergy} (Kirkpatrick-Nawaz \cite{KN})
For dimension $ N $, the free energy $\varphi$ has the formula:
\[ \varphi_N(\beta) = \begin{cases} 0, \quad \quad \quad \quad \quad \quad \quad \quad \text{ if } \beta <N , \\ \Phi_{\beta,N}(g^{-1}(\beta)),  \quad \quad \quad \;\, \text{ if } \beta\ge N, \end{cases}\] 
where $g^{-1}(\beta) = r$ with $$g(r) = g_N(r) := r \frac{ I_{\frac{N}{2}-1}(r) }{ I_\frac{N}{2}(r)},$$ and 
\[\Phi_{\beta,N}(r)= r \frac {I_\frac{N}{2}(r)}{I_{\frac{N}{2}-1}(r)}+\log\left[\frac{A_{N}}{A_{N-1}} \frac {r^{\frac{N}{2}-1}}{ B_N \pi I_{\frac{N}{2}-1}(r) }\right]-\frac{\beta}{2}\left(\frac {I_\frac{N}{2}(r)}{I_{\frac{N}{2}-1}(r)}\right)^2 ,\]
with
\[A_N := \frac{2  \pi^\frac{N}{2}}{\Gamma{\left(\frac{N}{2}\right)}}\] and 
\[B_N= \begin{cases} \prod_{k=0}^{\frac{N}{2}-1} |2k-1| , \quad \quad \quad \quad \quad \quad \quad \quad \text{ if $N$ even}, \\ \\ \frac{2^{\frac{N}{2}-1} \Gamma \left( \frac{N-1}{2}\right)}{\sqrt{\pi}},  \quad \quad \quad \quad\quad ~\quad \quad \quad \;\, \text{ if $N$ odd}. \end{cases}\]
 In particular, $\varphi$ and $\varphi'$ are continuous at  the critical threshold $\beta =N$, implying that the phase transition is second-order or continuous.
 
\end{thm}

\subsection{The critical density function in $O(N)$ models}

The limiting density for the critical case uses the (hyper-)spherical symmetry of the total spin for $O(N)$ models, giving the following non-normal limit theorem.

\begin{thm}\label{density_func} (Kirkpatrick-Nawaz \cite{KN})
At the critical temperature $\beta = N$, the random variable $W_n = \frac{c_N |S_n|^2}{n^{3/2}}$ has as its limit as $n \to \infty$ the random variable $X$ with density

\[ p_N(t) = \begin{cases} \frac{1}{Z} t^\frac{N-2}{2}~e^{-k t^2}, \text{ if } t \ge 0 , \\ 0, ~\quad \quad \quad \quad\; \text{ if } t <0, \end{cases}\]  
where $k = \frac {1} {N^2(4 N+8)}$ and $Z$ is the normalizing constant. To be precise about the rate of convergence, there exists a universal constant $C$ such that 
\[\sup_{\substack{\|h\|_\infty\le 1, \,\|h'\|_\infty\le 1\\\|h''\|_\infty\le
1}}\big|\E h(W_n)-\E h(X)\big|\le\frac{C\log(n)}{\sqrt{n}}.\]
\end{thm}

\begin{figure}[h]
\centering
\includegraphics[width=5in]{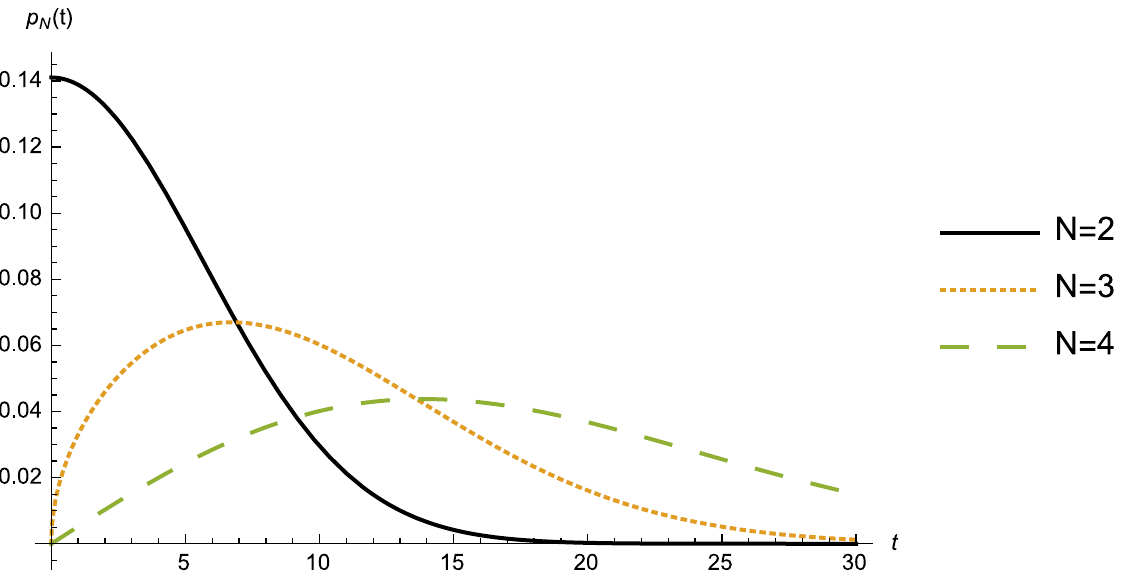}
\caption{Mean-field critical density functions $p_N$ for $2 \le N \le 4$ and $t \ge 0$. For the XY model $p_2(t)  = \frac{e^{-t^2/64}}{4 \sqrt{\pi}}$, for the Heisenberg model $p_3(t)  = \frac{\sqrt{t} e^{-t^2/180}}{5^{3/4} \sqrt{54} \Gamma[3/4]}$, and for the Toy model of the Higgs sector, $p_4(t)  = \frac{t e^{-t^2/384}}{192}$. }
\label{fig:5}
\end{figure}

The proof of this theorem is in \cite{KN} and includes methods from \cite{RECN,ENR,KM}. 

\end{document}